\documentclass{mn2e}
\usepackage{times}
\input{psfig.sty}

\begin{document}

%\draft

%%%%% AUTHORS -- PLACE YOUR OWN MACROS HERE %%%%%

\def\pim{$\pm$} 
\def\be{\begin{equation}}
\def\ee{\end{equation}}
\def\mdot{$\dot{m}$ }
\def\mpc{\,{\rm {Mpc}}}
\def\kpc{\,{\rm {kpc}}}
\def\kms{\,{\rm {km\, s^{-1}}}}
\def\msun{{$\rm M_{\odot}$}}
\def\Gyr{{\,\rm Gyr}}
\def\erg{{\rm erg}}
\def\sr{{\rm sr}}
\def\hz{{\rm Hz}}
\def\cm{{\rm cm}}
\def\sec{{\rm s}}
\def\eV{{\rm \ eV}}
\def\ledd{$L_{Edd}$~}
\def\mic{$\mu$ }
\def\ang{\AA }  % Angstrom
\def\cm2{cm$^2$ }
\def\se1{s$^{-1}$ }

\def\arcmin{\hbox{$^\prime$}}
\def\arcsec{\hbox{$^{\prime\prime}$}}
\def\degree{$^{\circ}$} 
\def\mic{$\mu$ }
\def\ang{\AA }  % Angstrom
\def\cm2{cm$^2$ }
\def\se1{s$^{-1}$ }

\def\gtsima{$\; \buildrel > \over \sim \;$}
\def\ltsima{$\; \buildrel < \over \sim \;$}
\def\prosima{$\; \buildrel \propto \over \sim \;$}
\def\gsim{\lower.5ex\hbox{\gtsima}}
\def\lsim{\lower.5ex\hbox{\ltsima}}
\def\simgt{\lower.5ex\hbox{\gtsima}}
\def\simlt{\lower.5ex\hbox{\ltsima}}
\def\simpr{\lower.5ex\hbox{\prosima}}
\def\la{\lsim}
\def\ga{\gsim}
\def\Lsun{\rm L_{\odot}}
\def\Rsun{\rm R_{\odot}} 
\def\sr{V404~Cyg~}
\def\gx{GX~339$-$4~}

\def\ie{{\frenchspacing\it i.e. }}
\def\eg{{\frenchspacing\it e.g. }}
\def\etal{{~et al.~}}

%================================================================================
\title[The radio spectrum of V 404 Cyg ]  
{The radio spectrum of a quiescent stellar mass black hole} 
\author[Gallo, Fender \& Hynes ]
{E. Gallo$^{1}$, R. P. Fender$^{1}$\thanks{Present address: School of Physics
and Astronomy, University of Southampton, Hampshire SO17 1BJ, United Kingdom},
R. I. Hynes$^{2}$\thanks{Present address: Louisiana State University,
Department of Physics and Astronomy, Baton Rouge, LA 70803-4001, USA}
\\ \\
$^{1}$ Astronomical Institute `Anton Pannekoek', University of
Amsterdam,\\$~$and 
Center for High Energy Astrophysics, Kruislaan 403, 1098 SJ Amsterdam, The Netherlands\\
$^{2}$ McDonald Observatory and Department of Astronomy,
The University of Texas at Austin, 1 University Station C1400, Austin, Texas
78712, USA \\}
\maketitle
%=============================================================================
\begin{abstract} 
Observations of \sr performed with the Westerbork Synthesis Radio Telescope at
four frequencies, over the interval 1.4$-$8.4 GHz, have provided us with the
first broadband radio spectrum of a `quiescent' stellar mass black hole.  The
measured mean flux density is of 0.35 mJy, with a spectral index $\alpha= +
0.09\pm0.19$ (such that $\rm S_{\nu}\propto \nu^{\alpha}$).  Synchrotron
emission from an inhomogeneous partially self-absorbed outflow of plasma
accounts for the flat/inverted radio spectrum, in analogy with hard state
black hole X-ray binaries, indicating that a steady jet is being produced
between a few $10^{-6}$ and a few per cent of the Eddington X-ray luminosity.
\end{abstract}  
%==================================================================================
\begin{keywords}
Binaries: general -- ISM: jets and outflows -- Individual: V404 Cyg 
\end{keywords}
%=================================================================================
\section{Introduction}

While accreting gas at relatively low rates, black hole candidates in X-ray
binary (BHXB) systems are able to power steady, collimated outflows of energy
and material, oriented roughly perpendicular to the orbital plane. The jet
interpretation of the radio emission from hard state BHXBs came before the
collimated structures were actually resolved with Very Long Based
Interferometry (VLBI) techniques.  In a seminal work, Blandford \& K\"onigl
(1979) proposed a model to interpret the flat radio spectrum of extragalactic
compact radio sources in terms of isothermal, conical outflows, or jets. A jet
model for X-ray binaries was later developed by Hjellming \& Johnston (1988),
in order to explain both the steady radio emission with flat/inverted spectra
observed in the hard state of BHXBs, and transient outbursts with optically
thin synchrotron spectra. We refer the reader to McClintock \& Remillard
(2004) and Fender (2004) for comprehensive reviews on X-ray states and radio
properties (respectively) of BHXBs.  High resolution maps of Cyg X-1 in the
hard X-ray state have confirmed the jet interpretation of the flat radio-mm
spectrum (Fender \etal 2001), imaging an extended, collimated structure on
milliarcsec scale (Stirling \etal 2001).  Further indications for the
existence of collimated outflows in the hard state of BHXBs come from the
stability in the orientation of the electric vector in the radio polarization
maps of GX~339$-$4 over a two year period (Corbel \etal 2000). This constant
position angle, being the same as the sky position angle of the large-scale,
optically thin radio jet powered by GX 339$-$4 after its 2002 outburst (Gallo
\etal 2004), clearly indicates a favoured ejection axis in the system.
Finally, the optically thick milliarcsec scale jet of the (somewhat peculiar)
BH candidate GRS 1915+105 (Dhawan, Mirabel \& Rodr\'\i guez 2000) in the
plateau state (Klein-Wolt
\etal 2002) supports the association of hard X-ray states of BHXBs with
steady, partially self-absorbed jets.

Having established this association, a natural question arises: what are the
required conditions for a steady jet to exist? We wonder especially whether
the jet survives in the very low luminosity, \emph{quiescent} X-ray
state. While radio emission from BHXBs in the thermal dominant (or high/soft)
state is suppressed up to a factor $\sim$50 with respect to the hard state
(\eg Fender \etal 1999; Corbel \etal 2001, and references therein), most
likely corresponding to the physical disappearance of the jet, little is known
about the radio behaviour of quiescent stellar mass BHs, mainly due to
sensitivity limitations. Among the very few systems detected in radio is V404
Cygni, which we shall briefly introduce in the next Section.

\subsection{V404 Cyg (=GS~2023+338)}

The X-ray binary system \sr is thought to host a strong BH candidate, with a
most probable mass of $\sim$ 12 \msun~(Shahbaz \etal 1994), and a low mass
K0IV companion star, with orbital period is of 6.5 days, and orbital
inclination to the line of sight is of about 56\degree~(Casares \& Charles
1994; Shahbaz 1994). Following the decay of the 1989 outburst that led to its
discovery (Makino 1989), the system entered a quiescent X-ray state, in which
it has remained ever since. The relatively high quiescent X-ray luminosity of
V404 Cyg (with an \emph{average} value of about $6 \times 10^{33}\times (\rm D/4 ~kpc)^2$ 
erg sec$^{-1}$ in the range $0.3-7.0$ keV; Garcia \etal 2001; Kong \etal 2002;
Hynes \etal 2004) is possibly related to the long orbital period and surely
indicates that the some accretion continues to take place at $\rm L_{X}\simeq
4\times 10^{-6}\rm L_{Edd}$, where $\rm L_{Edd}$ is the Eddington X-ray
luminosity (for a 12 \msun~BH). 

As reported by Hjellming \etal (2000), since (at least) early 1999 the system
has been associated with a variable radio source with flux density ranging
from 0.1 to 0.8 mJy on time scales of days and it is known to vary at optical
(Wagner \etal 1992; Casares
\etal 1993; Pavlenko \etal 1996; Hynes \etal 2002; Shahbaz \etal 2003; Zurita
\etal 2003) 
and X-ray wavelengths (Wagner \etal 1994; Kong \etal 2002; Hynes \etal 2004,
for a coordinated variability study) as well.  Yet no \emph{broadband} radio
spectrum of V404~Cyg in quiescence, nor of any other stellar mass BH below
$10^{-5} \rm L_{Edd}$, is available in the literature to date (see Corbel
\etal 2000 for a 2-frequency radio spectrum of GX~339-4 at $\sim$$ 10^{-5}\rm
L_{Edd}$).  Given the quite large degree of uncertainty about the overall
structure of the accretion flow in quiescence (e.g. Narayan, Mahadevan \&
Quataert 1999 for a review), it has even been speculated that the total power
output of a quiescent BH could be dominated by a radiatively inefficient
outflow (Fender, Gallo \& Jonker 2003) rather than by the local dissipation of
gravitational energy in the accretion flow. It is therefore of primary
importance to establish the nature of radio emission from quiescent BHXBs. In
this brief paper we show that the radio properties of \sr closely resemble
those of a canonical hard state BH, suggesting that there is no fundamental
difference in terms of radio behaviour between the quiescent and the canonical
hard X-ray state.  A comprehensive study of the spectral energy distribution
of \sr in quiescence, from radio to X-rays, will be presented elsewhere (Hynes
et al., in preparation).

\section{Radio emission from V404 Cyg}
\subsection{WSRT observations}

The Westerbork Synthesis Radio Telescope (WSRT) is an aperture synthesis
interferometer that consists of a linear array of 14 dish-shaped antennae
arranged on a 2.7 km East-West line. \sr was observed by the WSRT at two
epochs:~~{\bf{i)}} on 2001 December 28, start time 05:28 UT (MJD 52271.3), at
4.9 GHz (6~cm) and 8.4 GHz (3~cm), for 8 hours at each frequency; observations
were performed with the (old) DCB backend, using 8 channels and 4
polarizations; ~~{\bf{ii)}} on 2002 December 29, start time 06:29 UT (MJD
52637.3), at 1.4 GHz (21~cm), 2.3 GHz (13~cm), 4.9 GHz (6~cm) and 8.4 GHz
(3~cm) for a total of 24 hours. Frequency switching between 8.4$-$2.3 GHz and
4.9$-$1.3 GHz was operated every 30 minutes over the two 12 hour runs,
resulting in $\sim$ 5.5 hour on the target and $\sim$ 0.5 hour on the
calibration sources (3C~286 and 3C~48) at each frequency.  During this set of
observations, the WSRT was equipped with the DZB backend using eight IVC
sub-bands of 20 MHz bandwidth, 64 channels and 4 polarizations. Seven out of
the eight sub-bands were employed to reconstruct the images, as the sub-band
IVC-IF6 failed to detect any signal other than noise over the whole 24 hour
period.  The telescope operated in its
\emph{max-short} configuration, particularly well suited for observations
shorter than a full 12 hour synthesis, and with a minimum baseline of 36~m
(see http://www.astron.nl/wsrt/wsrtGuide/ for further details).  The data
reduction, consisting of editing, calibrating and Fourier transforming the
$(u,v)$-data on the image plane, has been performed with the MIRIAD
(Multichannel Image Reconstruction Image Analysis and Display) software (Sault
\& Killeen 1998). The 1.4 and 2.3 GHz data, containing several sources with
flux density well above 100 mJy, were self-calibrated in phase.
\subsection{Results: spectrum and variability}
\subsubsection{2001 December 28~~(MJD 52271.3)}

\begin{figure}
%\hspace{0.5cm}
\centering{\psfig{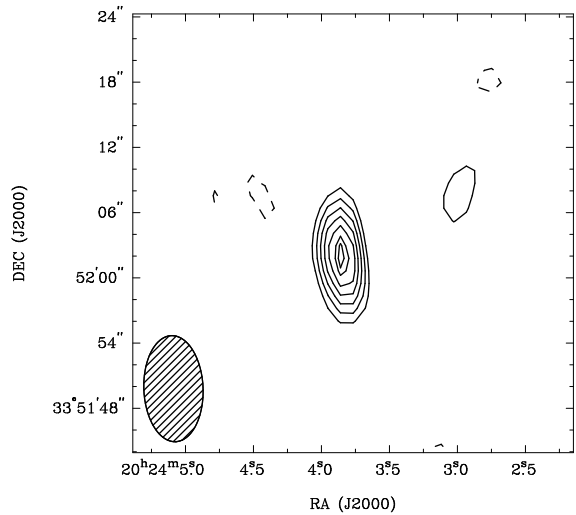}}
\caption{\label{map4.9}
Naturally weighted contour map of \sr as observed by Westerbork at 4.9~GHz on
2002 December 29 (MJD 52637.3); contour levels are at $-$3,3,4,5,6,7,8
times the rms noise level of 0.05 mJy; the synthesized beam is shown on the
bottom left corner. }
\end{figure}

\begin{table}
\caption{\label{table1} WSRT observations of \sr.}
\centering
\begin{tabular}{lccc}
\hline

date 		& $\nu$ (GHz) 	& $S_{\nu}$ (mJy) & $S/N$ \\
&&\\
2001-12-28	& 4.9		& 0.49 \pim 0.04  &12.2\\
\emph{(MJD 52271.3)}		&8.4		& 0.50 \pim 0.20  &2.5	\\
&&&\\
2002-12-29	&1.4 		& 0.34 \pim 0.08   & 4.2\\ 
\emph{(MJD 52637.3)}&2.4 		& 0.33 \pim 0.07  & 4.7\\ 
		&4.9 		& 0.38 \pim 0.05  & 7.6\\ 
		&8.4 		& 0.36 \pim 0.15  & 2.4\\ 
\hline
\end{tabular}
\flushleft
\end{table}
An unresolved (beam size of $\sim 5.8 \times 3.0$~arcsec$^{2}$ at 8.4 GHz)
$\sim$0.50 mJy radio source is detected at both 4.9 and 8.4 GHz, at the
position consistent with that of \sr ($\alpha$(J2000) = 20:24:03.78;
$\delta$(J2000) = +33:52:03.2; e.g. Downes \etal 2001). Table 1 lists the
measured flux densities with errors at each frequency; the corresponding
spectral index (hereafter defined as $ \alpha =\Delta~\rm log~S_{\nu} /
\Delta~\rm log~\nu $, such that $\rm S_{\nu}\propto \nu^{\alpha } $) is of
0.04\pim0.68; such a large error bar is mainly due to the high noise in the
8.4 GHz map (see Table 1). 
The signal/noise ratios are too low to measure 
linearly polarized flux from the source at the expected level of a few per
cent, assuming a synchrotron origin for the radio emission (see Section 3).
\subsubsection{2002 December 29~~(MJD 52637.3)}
\begin{figure}
\hspace{0.5cm}
\vspace{-0.3cm}
\centering{\psfig{figure=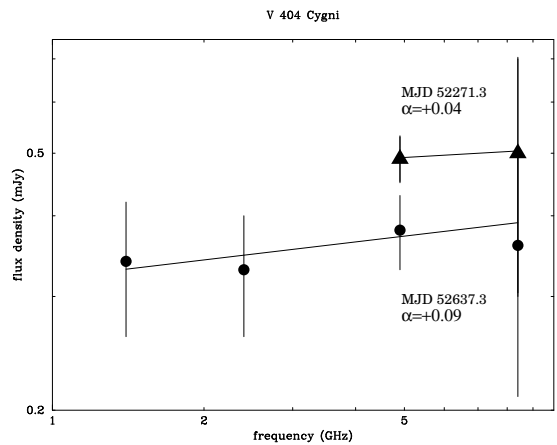,width=9cm,angle=-90}}
\vspace{-0.4cm}
\caption{\label{spectrum}  The radio spectrum of \sr as measured by the WSRT on 2001 December 28 (MJD
52271.3) and 2002 December 29 (MJD 52637.3); flux densities are listed in
Table 1.  }
\end{figure}

\sr is detected at four frequencies with a mean flux density of 0.35 mJy;
flux densities at each frequency are listed in Table 1. The fitted
four-frequency spectral index is $\alpha=0.09\pm0.19$.  Radio contours as
measured at 4.9~GHz are plotted in Figure 1, while Figure
\ref{spectrum} shows the radio spectra of \sr at two epochs. 

Since returning to quiescence, \sr is known to vary on time scales of days, or
even shorter, both in radio and in X-rays; such variability is actually
detected in our 2002 WSRT observations. The low flux of \sr makes it
practically impossible to subtract from the $(u,v)$-data all the other radio
sources in the field and generate a reliable light curve of the target. We
thus divided each of the two $\sim$ 11-hour data sets on-source in time
intervals of $\sim$ 5.5 hours (of which only $\sim$ 2.75 hours on source per
frequency, due to the frequency switching) and made maps of each time
interval. Significant variability (checked against other bright sources in the
field) is detected at 4.9 GHz: the flux density varied from 0.27\pim 0.07 mJy
in the first half of the observation, to 0.47\pim 0.07 in the second half.

\section{Discussion}

As mentioned in the introduction, synchrotron radiation from a relativistic
outflow accounts for the observed flat radio spectra of \emph{hard} state
BHXBs; we refer the interested reader to more thorough discussions in
e.g. Hjellming \& Han (1995), Mirabel \& Rodr\'\i guez (1999) and Fender
(2001; 2004).  Here we shall note that the \emph{collimated} nature of these
outflows is more debated, as it requires direct imaging to be
proven. Even though confirmations come from Very Long Based Array (VLBA)
observations of Cyg X-1 (Stirling \etal 2001) and GRS 1915+105 (Dhawan \etal
2000; Fuchs \etal 2003) in hard states, failure to image a collimated
structure in the hard state of XTE J1118+480 down to a synthesized beam of
0.6$\times$1.0 mas$^{2}$ at 8.4 GHz (Mirabel
\etal 2001) may challenge the jet interpretation (Fender \etal 2001). However,
apart from GRS 1915+105, which is persistently close to the Eddington rate
(see Fender \& Belloni 2004 for a review), Cyg X-1 in the hard state displays
a 0.1-200 keV luminosity of 2 per cent $\rm L_{Edd}$ (Di Salvo
\etal 2001), while XTE J1118+408 was observed at roughly one order of
magnitude lower level (\eg Esin \etal 2001). If the jet size scaled as the
radiated power, we would expect the jet of XTE J1118+408 to be roughly ten
times smaller than that of Cyg X-1 (which is 2$\times$6 mas$^{2}$ at 9 GHz, at
about the same distance), and thus still point-like in the VLBA maps presented
by Mirabel \etal (2001).

Garcia \etal (2003) have pointed out that long period ($\simgt 1$ day) BHXBs
undergoing outbursts tend to be associated with spatially resolved optically
thin radio ejections, while short period systems would be associated with
unresolved, and hence physically smaller, radio ejections. If a common
production mechanism is at work in optically thick and optically thin BHXB
jets (Fender, Belloni \& Gallo 2004), the above arguments should
apply to steady optically thick jets as well, providing an alternative
explanation to the unresolved radio emission of XTE J1118+480, with its 4 hour
orbital period, the shortest known for a BHXB. It is worth mentioning that, by
analogy, a long period system, like \sr, might be expected to have a
relatively larger optically thick jet.

\subsection{A synchrotron emitting outflow in the
quiescent state of V404~Cyg}

\subsubsection{Emission mechanism}

The WSRT observations of \sr performed on 2002 December 29 provide us with the
first broadband (1.4$-$8.4 GHz) radio spectrum of a stellar mass BH
candidate below $10^{-5}\rm L_{Edd}$. As we do not have direct
evidence (no linear polarization measurement, no especially high brightness
temperature, see below) for the synchrotron origin of the radio emission from
\sr in quiescence, we must first briefly explore different mechanisms, such as
free-free emission from an ionised plasma.
The donor in \sr is a K0IV star with most probable mass of 0.7 \msun~and
temperature around 4300 K (Casares \& Charles 1994; Shahbaz \etal 1994),
simply too cool to produce any observable free-free radio emission (see Wright
\& Barlow 1975). 
Alternatively, the accretion flow onto the compact object may provide the
needed mass loss rates and temperatures in order to produce a flat/inverted
free-free radio spectrum. In (line- and radiation-driven) disc wind models,
global properties such as the total mass loss rate and wind terminal velocity
depend mainly on the system luminosity (see \eg Proga \& Kallman 2002; Proga,
Stone
\& Drew 1998 and references therein); very high accretion rates are required 
in order to sustain significant mass loss rates and hence observable wind
emission, ruling out a disc wind origin for the observed radio flux from
V404~Cyg.  However, that mass loss via winds in sub-Eddington, radiatively
inefficient accretion flows (ADAFs) may be both dynamically crucial and quite
substantial, has been pointed out by Blandford \& Begelman (1999). Quataert \&
Narayan (1999) calculated the spectra of such advection dominated inflows
taking into account wind losses, and found that the observations of three
quiescent black holes, including V404 Cyg, are actually consistent with at
least 90 per cent of the mass originating at large radii to be lost to a wind.
Under the rough assumption that models developed for ionised stellar winds
(\eg Wright \& Barlow 1975, Reynolds 1986; see Dhawan \etal 2000 for an
application to the steady jet of GRS 1915+105) might provide an order of
magnitude estimate of the mass loss rate even for such `advection-driven'
winds, still the required mass loss rate in order to sustain the observed
radio emission for a
\emph{fully ionised} hydrogen plasma is close the Eddington accretion rate for
a 12 \msun~BH (assuming a 10 per cent efficiency in converting mass into
light). Lower ionisation parameters would further increase the needed mass
loss, bringing it to super-Eddington rates.  Even taking into account
geometrical effects, such as wind collimation and/or clumpiness, the
required mass loss rates can not be more than three orders of magnitude below
the spherical homogeneous wind, \ie still far too high for a $\simlt 10^{-5}$
Eddington BH to produce any observable radio emission.  As free-free emission
does not appear to be a viable alternative, we are led to the conclusion that
the radio spectrum of \sr in quiescence is likely to be synchrotron in
origin.  This conclusion is supported by polarization measurements during the
second phase (following a bright optically thin event) of the 1989 radio
outburst of
\sr, when a slow-decay, optically thick component
had developed (Han \& Hjellming 1992). At this time, after 1989 June 1-3, \sr
displayed the same flat/inverted spectrum of the present 2002 WSRT
observations, but was at a few mJy level, still high enough to allow the
detection of linearly polarized flux, which confirmed the synchrotron nature of
the emission. In addition, the roughly constant and similar polarization angles
measured at that time, indicated that the averaged magnetic field orientation
changed very little, if at all. As \sr entered a quiescent regime following
the decay of that outburst (it reached the typical quiescent flux densities
about 1 year after the outburst peak), it seems reasonable to assume that the
present $\sim$ 0.5 mJy radio emission with flat/inverted spectrum is of the
same nature as the few mJy flat/inverted spectrum component detected in 1989,
and therefore synchrotron in origin.

A further argument for the \emph{jet} interpretation of this synchrotron
emission is the fact that the radio and X-ray fluxes of \sr over the decline of
its 1989 outburst \emph{and} at its current quiescent X-ray and radio
luminosities, display the same non-linear correlation found to hold in
the whole hard state of BHXBs (Corbel
\etal 2003; Gallo, Fender \& Pooley 2003) and later extended to super-massive
nuclei in active nuclei as well (Merloni, Heinz \& Di Matteo 2003; Falcke,
K\"ording \& Markoff 2004), where there is little doubt about the jet origin
of the radio emission.

\begin{table}
\caption{Constraints on the size $\rm L$ of the radio emission region
in \sr; a distance of 4 kpc is adopted (Jonker \& Nelemans 2004).  }
\centering
\begin{tabular}{ccccc}
\hline
requirement	& frequency	&	&Size	&	      \\
		&    GHz           &cm		&$\rm R_{\odot}$ & mas \\
\\
${\rm T_{b}}< 10^{12}$K & 1.4&$\simgt 5 \times 10^{11}$ & $\simgt 7 $ & $\simgt
0.01$\\ ${\rm L}< c~\Delta t$ &4.9& $\simlt 6 \times
10^{14}$ & $\simlt 8530 $& $\simlt 10$\\
\hline
\end{tabular}
\flushleft
\end{table}

\subsubsection{Angular size}

The maximum brightness temperature for a galactic incoherent synchrotron
source is a (weak) function of the measured spectral index, the upper
frequency $\nu_{\rm up}$ of the synchrotron spectrum and the Doppler boosting
factor $D$ (e.g. Hughes \& Miller 1991).  For $\alpha=0.1$ and $\nu_{\rm
up}$=8.4 GHz, as in the case presented here, $T_{\rm b} \simlt 10^{12}\times
D^{1.2}$ K at 1.4 GHz.  With an orbital inclination of 56$^{\circ}$, the
Doppler boosting factor is likely to vary in the range $D=$1--1.1, calculated
for bulk Lorentz factors between 1--1.7. Assuming a distance to V404 Cyg of 4
kpc (Jonker \& Nelemans 2004 and references therein), we can thus derive a
minimum linear size for the (1.4 GHz) emitting region of $\sim 5 \times
10^{11}$ cm, or $\sim 7~\rm R_{\odot}$. This corresponds to about one fifth of
the system orbital separation (Shahbaz \etal 1994). For comparison, the
highest {\it measured} brightness temperature in a Galactic binary system is
of a few $10^{11}$ K at 5 GHz, during a flaring event in Cyg X-3 (Ogley \etal
2001).

Because of limits on the signal propagation speed, the 5.5-hour time scale
variability detected at 4.9 GHz gives an upper limit to the linear size $\rm
L$ of the variable region: $\rm L < 5.9 \times 10^{14}$ cm, or about 8530 $\rm
R_{\odot}$. At a distance of 4 kpc, this translates into an angular extent
$\theta \simlt 10$ mas at 4.9 GHz (see Table 2).
In the framework of standard conical jet models (Blandford \& K\"onigl 1979;
Hjellming \& Johnston 1988; Falcke \& Biermann 1996), flux variability could
be induced by e.g. the propagation of shocks within the compact outflow. These
shocks will not be visible until they reach the point along the outflow where
it becomes optically thin at the observing frequency.  The actual morphology
of the radio source will depend on the ratio between the thickness $\Delta r$
of the region where the variability occurs (the lower the observing
frequency, the higher the thickness) and its distance $\rm R$ from the core.
If ${\rm R} \gg \Delta r$, we would expect a double radio source with flux
ratios depending on Doppler boosting, while if ${\rm R}\simeq \Delta r$, then
we would expect to observe a continuous elongated structure.  The ratio $\Delta
r/\rm R$ is unknown in the case of \sr and could only be determined by
measuring \emph{delays} between different frequencies. For comparison, the
average flux rise time in the oscillations of the flat spectrum radio
component in GRS 1915+105 is of about a few minutes, while the infrared-radio
delays are typically of 15 min, indicating that the variable radio source
should not be too distant from the core (Mirabel \etal 1998; Fender \etal
2002, and references therein). Combined with the extended core morphology of
both Cyg X-1 (Stirling \etal 2001) and the core of GRS 1915+105 (Dhawan \etal
2000; Fuchs \etal 2003), this suggests that ${\rm R}\simeq \Delta r$ in these
two sources, but it is not clear how the outflow properties might scale with
the luminosity.  

Han \& Hjellming (1992), based on the same arguments,
were able to constrain the linear size of the optically \emph{thin} ejection
associated with the fast-decay phase in the radio light curve of the 1989
outburst: from the 5~min time scale variability measured on 1989 June 1st, and
from brightness temperature limits, they derived $\theta\simeq 0.2$ mas.
Fluctuations on time scales of tens of minutes were later measured during the
slow-decay phase, when an optically thick component had developed, and
interpreted as possible hot shocks propagating downstream an underlying
compact jet. By analogy, this would appear a reasonable explanation for the
5-hour time scale variability detected in our WSRT observations as well.

\section{Summary}

WSRT observations of \sr performed on 2002 December 29 (MJD 52637.3) at four
frequencies over the interval 1.4$-$8.4 GHz have provided us with the first
broadband radio spectrum of a quiescent (with average $\rm L_X$ of a few
$10^{-6}\rm L_{Edd}$) stellar mass BHXB. We measured a mean flux density of
0.35 mJy, and a flat/inverted spectral index $\alpha= 0.09\pm0.19$. WSRT
observations performed one year earlier, at 4.9 and 8.4 GHz, resulted in a
mean flux density of 0.5 mJy, confirming the relatively stable level of radio
emission from \sr on a year time-scale; even though the spectral index was not
well constrained at that time, the measured value was consistent with the
later one. 

Synchrotron emission from an inhomogeneous, optically thick
relativistic outflow of plasma seems to be the most likely explanation for the
flat radio spectrum, in analogy with hard state BHXBs (Fender 2001).
Optically thin free-free emission as an alternative explanation is ruled out
on the basis that mass loss rates far too high would be required, either from
the companion star or from the inflow of plasma to the accretor. The
collimated nature of this outflow remains to be proven; based on brightness
temperature arguments and the 5.5-hour time-scale variability detected at 4.9
GHz, we conclude that the angular extent of the radio source is constrained
between 0.01 at 1.4 GHz and 10 mas at 4.9 GHz (at a distance of 4 kpc; Jonker
\& Nelemans 2004). In the context of standard self-absorbed jet models, the
flux variability may be due to shocks or clouds propagating in an
inhomogeneous jet.

If our interpretation is correct, a compact steady jet is being produced by
BHXBs between a few $10^{-6}$ and $\sim$$ 10^{-2}$ times the Eddington
luminosity, supporting the notion of quiescence as a low luminosity level of
the standard hard state. However, as \sr is the most luminous quiescent BHXB
known to date, the existence of a steady jet in this system does not
automatically extend to the whole quiescent state of stellar mass
BHs. Sensitive radio observations of the nearby, truly quiescent system
A0620$-$00 (three orders of magnitude less luminous than \sr in X-rays;
e.g. Kong
\etal 2002), will hopefully provide an answer about the ubiquity of compact
jets from stellar mass black holes with a hard spectrum.

%=================================================================
\section*{Acknowledgments}
EG wishes to thank Raffaella Morganti for her kind assistance in the data
reduction, and Alex de Koter for suggestions.~RIH is supported by NASA
through Hubble Fellowship grant $\#$ HF-01150.01-A awarded by STScI, which is
operated by AURA, for NASA, under contract NAS 5-26555.  The Westerbork
Synthesis Radio Telescope is operated by the ASTRON (Netherlands Foundation
for Research in Astronomy) with support from the Netherlands Foundation for
Scientific Research NWO.
%=================================================================

\end{document}